\documentclass[10pt]{article}

\usepackage{amsmath}
\usepackage{amssymb}

\usepackage{graphicx}

\usepackage{cite}

\usepackage{color} 

\usepackage[labelfont=bf,labelsep=period,justification=raggedright]{caption}

\makeatletter
\renewcommand{\@biblabel}[1]{\quad#1.}
\makeatother

\date{}

\begin{document}

\begin{flushleft}
{\Large
\textbf{A Scientometric Prediction of the Discovery of the First Potentially Habitable Planet with a Mass Similar to Earth}
}
\\
Samuel Arbesman$^{1,2,\ast}$, 
Gregory Laughlin$^{3}$
\\
\bf{1} Department of Health Care Policy, Harvard Medical School, Boston, MA, USA
\\
\bf{2} Institute for Quantitative Social Science, Harvard University, Cambridge, MA, USA
\\
\bf{3} Department of Astronomy and Astrophysics, University of California, Santa Cruz, CA, USA
\\
$\ast$ E-mail: arbesman@hcp.med.harvard.edu
\end{flushleft}

\section*{Abstract}

{\it Background: }
The search for a habitable extrasolar planet has long interested scientists, but only recently have the tools become available to search for such planets. In the past decades, the number of known extrasolar planets has ballooned into the hundreds, and with it the expectation that the discovery of the first Earth-like extrasolar planet is not far off.

{\it Methodology/Principal Findings: }
Here we develop a novel metric of habitability for discovered planets, and use this to arrive at a prediction for when the first habitable planet will be discovered. Using a bootstrap analysis of currently discovered exoplanets, we predict the discovery of the first Earth-like planet to be announced in the first half of 2011, with the likeliest date being early May 2011.

{\it Conclusions/Significance: }
Our predictions, using only the properties of previously discovered exoplanets, accord well with external estimates for the discovery of the first potentially habitable extrasolar planet, and highlights the the usefulness of {\it predictive} scientometric techniques to understand the pace of scientific discovery in many fields.


\section*{Introduction}

The search for a habitable extrasolar planet has long interested scientists, but only recently have the observational tools become available to search for such planets \cite{Santos}. Beginning in 1995 with the discovery of an extrasolar planet around Pegasi 51, a star much like our own \cite{MayorQueloz95}, the number of confirmed extrasolar planets has expanded into the hundreds. We now have a panoply of physical and orbital data on planets outside the solar system, and with it an increase in understanding of the formation of planetary systems. However, the holy grail of extrasolar planetary research -- an Earth-like planet -- has yet to be discovered. This search has more recently become even more intense with such large-scale surveys as NASA's Kepler mission \cite{Kochetal2010}.

While many, astronomers included, have speculated about when the first habitable planet might be discovered \cite{decisionPoll}, no quantitative scientometric analysis has been performed. In order to do this, we develop a metric of habitability for all discovered planets, and use this to arrive at a prediction for when the first habitable planet is expected to be discovered.

Of course, predicting future scientific and technological progress is a slippery and difficult process. While there have been many such successes, such as Moore's Law \cite{Moore}, history is littered with predictions that are far off the mark \cite{Grossman}.

Here too there are many difficulties. Estimating the habitability of planets is itself a complicated process with many parameters \cite{Vogel}, but most research dwells on the combination of two properties of a planet: surface temperature necessary for liquid water, and planetary mass \cite{Schilling}. Using these guidelines, we constructed a simple habitability metric. Using the habitability time series of previously discovered exoplanets, we created a bootstrap method to predict when the first Earth-like planet would be discovered. We predict the announcement of its discovery with a high probability by mid-year 2011.

\section*{Materials and Methods}

\subsection*{Habitability Metric}

Using the calculated mass and temperature of a planet, we constructed a simple habitability metric,$H_k$, for a given planet $k$, where $0$ is uninhabitable, and $1$ is an Earth-like planet. $H_k$ is defined as the following:

\begin{equation}
H_k = H^{T(a)}_k H^{T(b)}_k H^M_k
\end{equation}

where $H_k$ is the product of three sub-measures, each themselves on a scale of 0 to 1:

\begin{eqnarray}
H^{T(a)}_k &=& \text{habitability of surface temperature at } a \\
H^{T(b)}_k &=& \text{habitability of surface temperature at } b \\
H^M_k &=& \text{habitability of planetary mass}
\end{eqnarray}

The formulas for the submeasures that makes up $H_k = H^{T(a)}_k H^{T(b)}_k H^M_k$ are below, where ${M_\oplus}$ is one Earth mass, $W^T$ is the approximate width of half the acceptable temperature range (here we used $75\,{\mathrm K}$), $T_0$ is the midpoint of the range ($T_0 = 323\,{\mathrm K}$), $W^M = 0.5$ (for $H^{T(x)}_k$, and $x$ can either be $a$ (semi-major axis) or $b$ (semi-minor axis) of planet $k$):

\begin{eqnarray}
H^{T(x)}_k &=& \frac{1}{1+e^{-(T-T_0)+W^{T}}}\frac{1}{1+e^{(T-T_0)+W^{T}}} {\left( 1+e^{W^{T}} \right)}^2 \\ [6pt]
{H^M_k} &=& \frac{1}{1+e^{-(\frac{M}{M_\oplus}-M_\oplus)+{W^M}}} \frac{1}{1+e^{(\frac{M}{M_\oplus}-M_\oplus)+{W^M}}} {\left( 1+e^{{W^M}} \right)}^2
\end{eqnarray}

Each sub-measure is simply the product of two opposing modified logistic curves with some rescaling, which yield functions of the forms seen in Fig.~1. Note that while there is a positive probability assigned to negative masses, this is simply a byproduct of the symmetric nature of the function, and does not affect the calculations, as there are no such planets with negative mass. For $H$ to be 1, all sub-measures must themselves be 1. Due to the ``step-function-like'' nature of these sub-measures, $H$ is likely to be either very low near 0 or very close to 1. Additional separate conditions (such as presence or absence of atmosphere) would make the model more restrictive and necessarily lower $H$, so our model is erring optimistically on habitability, is an upper bound on $H$.

\begin{figure}[!ht]
\begin{center}
\includegraphics[width=240pt]{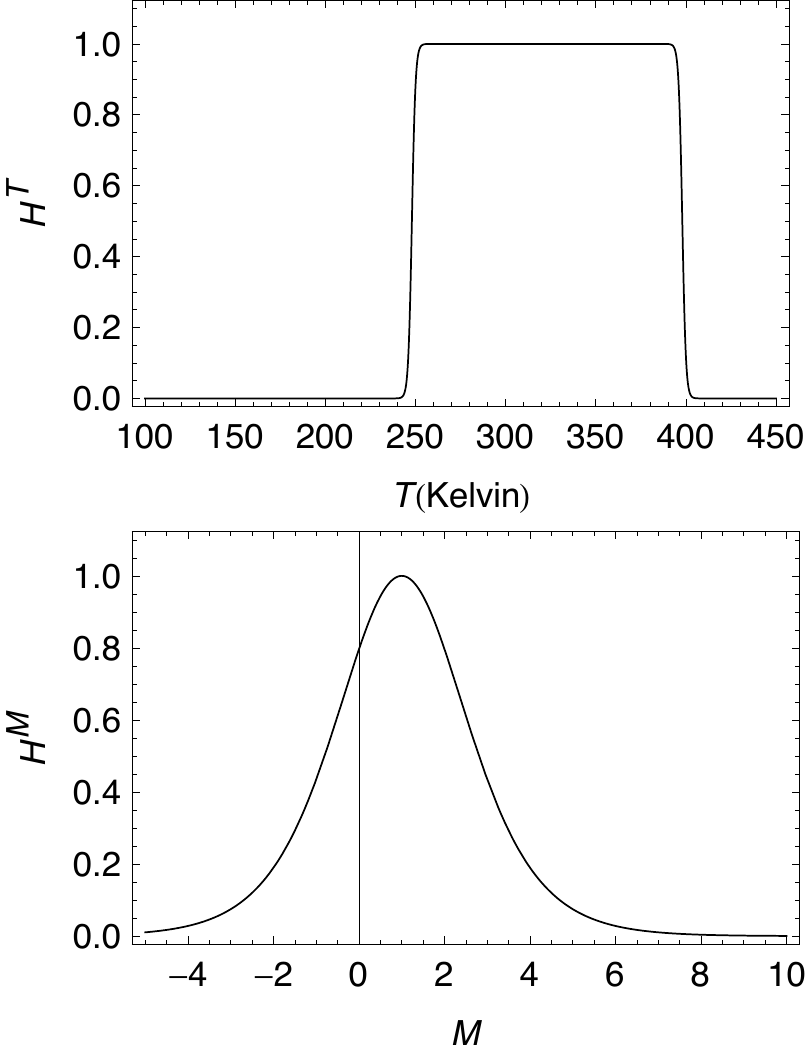}
\end{center}
\caption{
{\bf Habitability metric curves.} A. Habitability metric for temperature ${H^{T(x)}}$ as temperature varies. B. Habitability metric for $H^M$ as mass, in Earths, varies.
}
\label{Figure_label}
\end{figure}

Note that while a nominal temperature, $T=400\,{\mathrm K}$ exceeds the boiling point of water, this value is representative only of the simple blackbody equilibrium temperature at the substellar point. Actual surface temperatures for potentially habitable planets will be controlled by a host of effects, some well-understood, some entirely speculative: The stellar flux is intercepted by an area $\pi R_{\mathrm pl}^{2}$, but must warm a planet of surface area $4\pi R_{\mathrm pl}^{2}$. Our assigned limits on potential habitability represent correspond to a habitable zone whose outer radius is a factor of 2.6 times larger than its inner radius. Given the uncertainties on what constitutes the range of potentially habitable environments, this ratio is intended to be somewhat optimistic. Estimates by Mischna et al. (2000) \cite{Mischna}, for example, advocate a ratio of 2.1, whereas the influential study of Kasting et al. (1993) \cite{Kasting} found a ratio of bounding distances equal to 1.76. The atmosphere may provide a significant greenhouse warming effect. The planet may have endogenous sources of energy, and cloud cover can provide a significant reflective albedo. For a recent detailed discussion of the geophysical and atmospheric factors relevant to potentially habitable planets, see, e.g. Lammer et al. (2010) \cite{Lammer}.

Calculations of planetary surface temperature involve assumptions of stars on the Main Sequence and black body radiation models, and were as follows \cite{Lammer}:

\begin{equation}
T_a = \left[ \frac{\left(\frac{M_o}{M}\right)^{3.5}\left(3.839 \times 10^{26} \text{W} \right)}{4\pi\sigma_{sb}\left[\left(\frac{R_o}{R}\right)\left(6.955 \times 10^8 \text{m}\right) \right]^2}\right]^{1/4}
\end{equation}

Using these metrics, $H_k$ was calculated from readily available data \cite{Exoplanets} for all 370 planets in the dataset.

\subsection*{Discovery Date Prediction}

In order to create a robust estimate of the date of discovery of the first Earth-like planet, the following factors were considered:

\begin{enumerate}
\item The extrasolar planets considered were detected by two methods: radial velocity (RV) and transit. The radial velocity method detects a planet using the Doppler effect to determine the motion of the planet's star, and the transit method detects a planet using changes in brightness of the star due to the planet's transit in front of it. While the transit method provides accurate estimates of the mass, $M$ of a planet, the radial velocity method yields an estimate for $M \sin(i)$, where $i$ is an unknown inclination for the planetary system, thereby only giving a lower bound for $M$.
\item Any estimate of the detection of the first Earth-like planet will necessarily be dependent on the vagaries of those planets which were previously discovered. If other stars were examined, a different set of of planets, and therefore $H$ values, would have been found.
\end{enumerate}

A bootstrap analysis was conducted, which accounts for the $M$ estimates for planets discovered by the radial velocity method, and provides a robust estimate of the date of discovery. Each realization consisted first of calculating $M$ for each RV-detected planet in the dataset by drawing an inclination randomly chosen from the surface of the sphere. We then sampled 370 planets, each with its year of discovery, from the complete planetary data set with replacement, in order to create a bootstrapped time series of $H$ values of exoplanet discovery. The date of discovery chosen for each extrasolar planet was the mid-point of the year of discovery (including for 2010), due to the lack of more precise data.

To predict when the first habitable planet ($H \approx 1$) will be discovered, we examined the upper envelope of each realization's $H$ values by year (the points described by the highest habitability metric for each year). Since the upper and lower bound of $H$ are known to be $1$ and $0$, respectively, fitting a logistic equation is appropriate, and is similar to many other discovery curves, such as the number of mammalian species or number of chemical elements \cite{Arbesman,Price}. A logistic best fit of of the upper envelope of $H$ over time, $H(t)$, where the parameters $R$ and $y$ were allowed to vary, is as follows:
 
\begin{equation}
H(t) =  \frac{1}{1+e^{-R(y+t)}}
\end{equation}

The logistic curves of best-fit were performed using non-linear least squares fits on the sampled $H$ values. Due to the step-like function of $H(t)$, the best fit curves were extremely sensitive to initial conditions. While parameters close to the final fits were chosen for precision, variation in the results can be introduced by choosing different initial conditions. To determine the date as a fraction of the year, the value at which the logistic function first reached $H(t)=0.999$ was calculated. To test the robustness of using this assumption, $H(t)=0.99$ was used, which predicted early April 2011, and $H(t)=0.9999$ which predicted early June 2011.

Other robustness checks were performed as well. For either a bounds of $W^T = 50 {\mathrm K}$ or $W^M = 0.5$ for the habitability equation, a prediction of discovery in early May 2010 was found. It is likely then that the precision of the fit, along with the data available, are what drive the prediction. Assuming an error of a month in either direction is therefore reasonable. In order to convert this to days and months of a year, non-leap years were assumed. 

\section*{Results and Discussion}

We examined 370 exoplanets, all of which have well-characterized properties. Doing so, yields $H=0$ for the majority of exoplanets. Notably, Gliese 581 d, thought to be in the habitable zone, yields the highest value, with about $H=0.01$, and is still quite low. While some authors (e.g. Wordsworth et al. 2010 \cite{Wordsworth}) actually argue that Gliese 581 d is potentially habitable, we are of the opinion that its measured $Msin(i)=7.1 M_{\oplus}$ leads to an expected mass close to 10 Earth masses, and a possibly water-dominated composition more akin to an ice giant planet such as Uranus or Neptune than to a terrestrial planet like the Earth.

We conducted 10,000 realizations of the bootstrap method, where the data could successfully be fit to a curve, in order to arrive at a distribution of dates of discovery of the first habitable planet. This distribution is heavy-tailed, as seen in Fig.~2, with a median date of discovery of early May 2011 (2011.34). Additionally, detecting an Earth-like planet by the end of 2013 has a $2/3$ probability, we reach 75\% in 2020, and don't achieve 95\% likelihood until 2264.

\begin{figure}[!ht]
\begin{center}
\includegraphics[width=240pt]{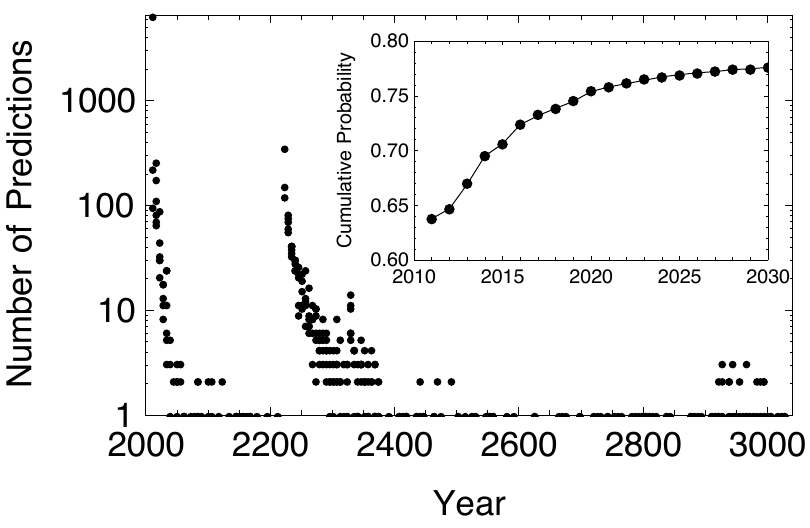}
\end{center}
\caption{
{\bf Distribution of 10,000 realizations of bootstrap analysis by year.} Inset shows the cumulative probability distribution of the year of discovery for 2011-2030.
}
\label{Figure_label}
\end{figure}

An example realization is shown in Fig.~3, where the best-fit logistic curve arrives at $H=1$ (the upper border), in the first half of the year 2011. More precisely, it reaches $H \approx 1$ at about one-third ($0.34$) through the year, which is early May 2011.

\begin{figure}[!ht]
\begin{center}
\includegraphics[width=240pt]{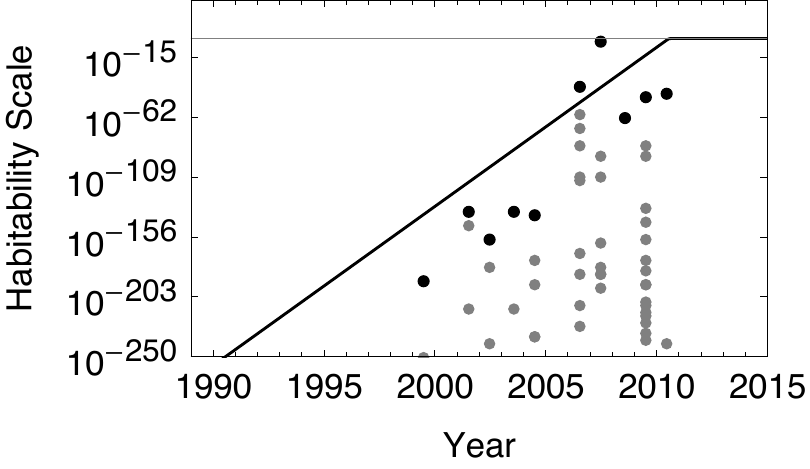}
\end{center}
\caption{
{\bf A single realization of the habitability of extrasolar planets over time.} $H$ values for the extrasolar planets are plotted, with those of the upper envelope (maximum $H$ for a given year of discovery) indicated in black. The black curve is the logistic best-fit curve of the upper envelope, using a nonlinear model, where $R=28.78$ and $y=2011.10$. The horizontal grey line indicates the maximum value of $H=1$, the presence of an Earth-like habitable planet.
}
\label{Figure_label}
\end{figure}

Additionally, we conducted the same bootstrap analysis for subsets of the planetary dataset up to the end of the years 2001-2010 (prior to this, there is not enough data to yield robust estimates). This allows us to determine what the likeliest date of the discovery of an Earth-like planet would have been predicted to be, if the analysis were conducted throughout the previous decade. The median dates of discovery are shown in Fig.~4.

\begin{figure}[!ht]
\begin{center}
\includegraphics[width=240pt]{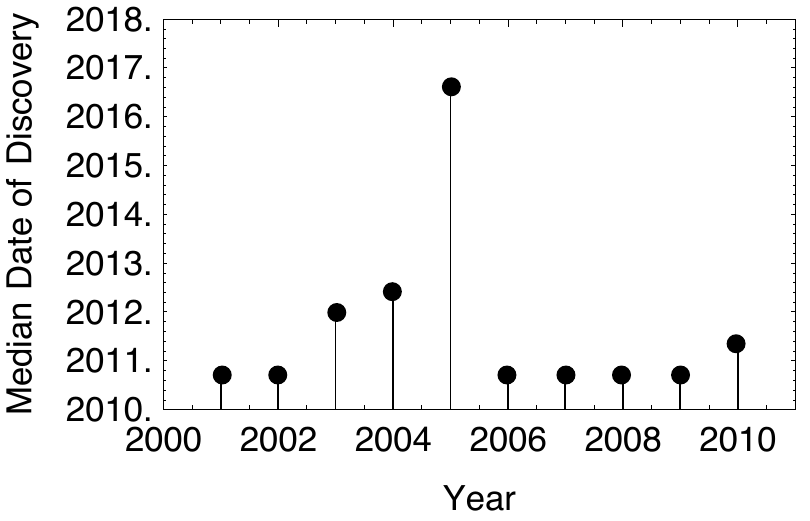}
\end{center}
\caption{
{\bf Median date of discovery using planetary data up to the end of a given year.} Results are from a bootstrap analysis for the years 2001-2010.
}
\label{Figure_label}
\end{figure}

The creation of a single metric of habitability, $H$, allows for quantitative prediction of when the first Earth-like planet is expected to be discovered -- in this case, a date of early May 2011. Of course, this prediction of when the discovery of the first Earth-like planet will be announced has ignored technological advancement entirely, as well as many other factors. However, technological progress can often be well-described by a functional form independent of the processes underlying its advancement \cite{Magee}. Similarly, it is likely that the multiple methods of extrasolar planet discovery (such as radial velocity and transit methods) combine to yield relatively smooth curves on the march towards further discovery. The testable prediction given here is likely found to be accurate in the coming months, given the recent launch and ongoing results of many projects.

A great deal of current interest is focused on NASA's ongoing Kepler mission \cite{Borucki}. The Kepler spacecraft employs the photometric transit method to detect planet candidates, and it is an open question as to whether this method can achieve the first detection of a planet with $H \approx 1$. While the initial results of Kepler were released on June 15, 2010, the Kepler team has delayed publication of 400 of the most promising extrasolar planetary candidates until February 2011. Within this large pool of withheld candidates, it is virtually certain that some have radii that are observationally indistinguishable from Earth's radius. It is likely, however, that because of the limited time base line of the mission to date, the Kepler planet candidates to published in February 2011 may be too hot to support significant values for $H$.

In order to determine how useful Kepler will be in the search for a habitable planet, we reran our prediction analysis using only those planets discovered using the transit method. And it turns out that the method is unable to converge on a likely date of discovery, due to the paucity of the data (62 planets) and the low $H$ values for these planets. No doubt Kepler will increase this number of planets, but this provides a counter-balance to the assumption that the Kepler team will discover the first habitable planet.

It must be noted that by publicizing our prediction, there is a concern that it will become accurate, simply due to the well-studied Hawthorne Effect \cite{Landsberger}. However, due to the large number of observations and long periods of time required to confirm an extrasolar planet discovery, it is unlikely that our prediction at this time will appreciably affect the announcement of the discovery of an Earth-like planet.

Therefore, it is reasonable to use the habitability metric curve as a rough prediction for when the first potentially habitable planet will be discovered, in this case, as early as May 2011, and likely by the end of 2013. 

\section*{Acknowledgments}

We would like to thank Jukka-Pekka Onnela for reading drafts of this manuscript and David Charbonneau for the initial conversation which prompted this investigation.

\bibliography{exoplanets.bib}

\begin{thebibliography}{10}
\providecommand{\url}[1]{\texttt{#1}}
\providecommand{\urlprefix}{URL }
\expandafter\ifx\csname urlstyle\endcsname\relax
  \providecommand{\doi}[1]{doi:\discretionary{}{}{}#1}\else
  \providecommand{\doi}{doi:\discretionary{}{}{}\begingroup
  \urlstyle{rm}\Url}\fi
\providecommand{\bibAnnoteFile}[1]{%
  \IfFileExists{#1}{\begin{quotation}\noindent\textsc{Key:} #1\\
  \textsc{Annotation:}\ \input{#1}\end{quotation}}{}}
\providecommand{\bibAnnote}[2]{%
  \begin{quotation}\noindent\textsc{Key:} #1\\
  \textsc{Annotation:}\ #2\end{quotation}}
\providecommand{\eprint}[2][]{\url{#2}}

\bibitem{Santos}
Santos N, Benz W, Mayor M (2005) Extrasolar planets: Constraints for planet
  formation models.
\newblock Science 310: 251-255.
\bibAnnoteFile{Santos}

\bibitem{MayorQueloz95}
Mayor M, Queloz D (1995) A jupiter-mass companion to a solar-type star.
\newblock Nature 378: 355-359.
\bibAnnoteFile{MayorQueloz95}

\bibitem{Kochetal2010}
{Koch} DG, {Borucki} WJ, {Basri} G, {Batalha} NM, {Brown} TM, et~al. (2010)
  Kepler mission design, realized photometric performance, and early science.
\newblock Astrophysical Journal Letters 713: L79-L86.
\bibAnnoteFile{Kochetal2010}

\bibitem{decisionPoll}
Laughlin G.
\newblock http://oklo.org/.
\bibAnnoteFile{decisionPoll}

\bibitem{Moore}
Moore G (1965) Cramming more components onto integrated circuits.
\newblock Electronics 38.
\bibAnnoteFile{Moore}

\bibitem{Grossman}
Grossman L (2004) Forward thinking.
\newblock TIME October 11, 2004.
\bibAnnoteFile{Grossman}

\bibitem{Vogel}
Vogel G (1999) Planetary systems: Expanding the habitable zone.
\newblock Science 286: 70-71.
\bibAnnoteFile{Vogel}

\bibitem{Schilling}
Schilling G (2007) Exoplanets: Habitable, but not much like home.
\newblock Science 316: 528b.
\bibAnnoteFile{Schilling}

\bibitem{Mischna}
{Mischna} MA, {Kasting} JF, {Pavlov} A, {Freedman} R (2000) {Influence of
  carbon dioxide clouds on early martian climate}.
\newblock Icarus 145: 546-554.
\bibAnnoteFile{Mischna}

\bibitem{Kasting}
{Kasting} JF, {Whitmire} DP, {Reynolds} RT (1993) {Habitable Zones around Main
  Sequence Stars}.
\newblock Icarus 101: 108-128.
\bibAnnoteFile{Kasting}

\bibitem{Lammer}
{L}ammer H, {S}elsis F, {C}hassefi{\`e}re E, {B}reuer D, {G}rie{\ss}meier JM,
  et~al. (2010) Geophysical and atmospheric evolution of habitable planets.
\newblock Astrobiology 1: 45-68.
\bibAnnoteFile{Lammer}

\bibitem{Exoplanets}
Survey CP (2010) Exoplanet Orbit Database.
\newblock http//www.exoplanets.org/: Accessed June 14, 2010.
\bibAnnoteFile{Exoplanets}

\bibitem{Arbesman}
Arbesman S (2010) Quantifying the ease of scientific discovery.
\newblock Scientometrics.
\bibAnnoteFile{Arbesman}

\bibitem{Price}
Price D (1986) Little science, big science-- and beyond.
\newblock New York: Columbia University Press.
\bibAnnoteFile{Price}

\bibitem{Wordsworth}
Wordsworth R, Forget F, Selsis F, Madeleine JB, Millour E, et~al. (2010) Is
  gliese 581d habitable? some constraints from radiative-convective climate
  modeling.
\newblock Technical Report arXiv:1005.5098.
\bibAnnoteFile{Wordsworth}

\bibitem{Magee}
Koh H, Magee C (2006) A functional approach for studying technological
  progress: Application to information technology.
\newblock Technological Forecasting and Social Change 73: 1061-1083.
\bibAnnoteFile{Magee}

\bibitem{Borucki}
Borucki WJ, Koch D, Basri G, Batalha N, Brown T, et~al. (2010) Kepler
  planet-detection mission: Introduction and first results.
\newblock Science 327: 977-980.
\bibAnnoteFile{Borucki}

\bibitem{Landsberger}
Landsberger H (1958) Hawthorne Revisited.
\newblock Ithaca.
\bibAnnoteFile{Landsberger}

\end{thebibliography}


\end{document}